# Self-regulation of human brain activity using simultaneous real-time fMRI and EEG neurofeedback

Vadim Zotev [a], Raquel Phillips [a], Han Yuan [a], Masaya Misaki [a], Jerzy Bodurka [a,b,*]

[a] Laureate Institute for Brain Research, Tulsa, OK, USA, [b] College of Engineering, University of Oklahoma, Tulsa, OK, USA

**Abstract**

Neurofeedback is a promising approach for non-invasive modulation of human brain activity with applications for treatment of mental disorders and enhancement of brain performance. Neurofeedback techniques are commonly based on either electroencephalography (EEG) or real-time functional magnetic resonance imaging (rtfMRI). Advances in simultaneous EEG-fMRI have made it possible to combine the two approaches. Here we report the first implementation of simultaneous multimodal rtfMRI and EEG neurofeedback (rtfMRI-EEG-nf). It is based on a novel system for real-time integration of simultaneous rtfMRI and EEG data streams. We applied the rtfMRI-EEG-nf to training of emotional self-regulation in healthy subjects performing a positive emotion induction task based on retrieval of happy autobiographical memories. The participants were able to simultaneously regulate their BOLD fMRI activation in the left amygdala and frontal EEG power asymmetry in the high-beta band using the rtfMRI-EEG-nf. Our proof-of-concept results demonstrate the feasibility of simultaneous self-regulation of both hemodynamic (rtfMRI) and electrophysiological (EEG) activities of the human brain. They suggest potential applications of rtfMRI-EEG-nf in the development of novel cognitive neuroscience research paradigms and enhanced cognitive therapeutic approaches for major neuropsychiatric disorders, particularly depression.

*Keywords:* Neurofeedback, real-time fMRI, EEG, EEG-fMRI, rtfMRI-EEG neurofeedback, emotion, amygdala, frontal EEG asymmetry

## Introduction

Neurofeedback is a general methodological approach that uses various neuroimaging techniques to acquire real-time measures of brain activity and enable volitional self-regulation of brain function. The development of real-time functional magnetic resonance imaging (rtfMRI) (Cox et al., 1995), in which fMRI data processing and display keep up with MRI image acquisition, has made it possible to implement rtfMRI neurofeedback (e.g. Weiskopf et al., 2004; deCharms, 2008; Sulzer et al., 2013). rtfMRI neurofeedback (rtfMRI-nf) allows a subject inside an MRI scanner to watch and self-regulate blood-oxygenation-level-dependent (BOLD) fMRI activity in target region(s) of his/her own brain in what is experienced as real time. Studies performed over the past decade demonstrated the feasibility of rtfMRI-nf-based self-regulation of various localized brain regions, including dorsal anterior cingulate cortex (Weiskopf et al., 2003), rostral anterior cingulate cortex (deCharms et al., 2005), auditory cortex (Yoo et al., 2006), anterior insular cortex (Caria et al., 2007; Ruiz et al., 2013), inferior frontal gyrus (Rota et al., 2009), supplementary motor area (Subramanian et al., 2011), subgenual anterior cingulate cortex (Hamilton et al., 2011), amygdala (Zotev et al., 2011), orbitofrontal cortex (Hampson et al., 2012), primary motor cortex (Berman et al., 2012), and others. Implementations of rtfMRI-nf for regulation of extended networks of brain areas defined using either functional localizers (e.g. Johnston et al., 2010; Linden et al., 2012) or support vector classification (LaConte, 2011; Sitaram et al., 2011) have also been reported.

In contrast to rtfMRI, which has temporal resolution equal to fMRI repetition time *TR* (order of a few seconds), electroencephalography (EEG) has millisecond temporal resolution and can record electrophysiological brain activity as it evolves in actual real time. EEG neurofeedback (EEG-nf) allows a subject to control certain characteristics of his/her own electrical brain activity as measured by EEG electrodes connected to the scalp. EEG-nf has a longer history and more reported applications to various patient populations than rtfMRI-nf. Some examples include: the sensorimotor rhythm (SMR) EEG-nf for treatment of epilepsy and seizure disorders (e.g. Sterman and Friar, 1972; Sterman, 2000); the SMR-theta and beta-theta EEG-nf for treatment of attention-deficit/hyperactivity disorder (e.g. Lubar and Lubar, 1984; Levesque et al., 2006; Gevensleben et al., 2009); the alpha-theta EEG-nf for treatment of substance use disorders (e.g. Peniston and Kulkosky, 1989; Sokhadze et al., 2008); the alpha-theta EEG-nf for deep relaxation (e.g. Egner et al., 2002) and creative performance enhancement (e.g. Gruzelier, 2009); the upper-alpha EEG-nf for cognitive enhancement (Hanslmayr et al., 2005; Zoefel et al., 2011); the frontal asymmetry EEG-nf for emotion regulation (Allen et al., 2001); and the high-beta EEG-nf for treatment of major depressive disorder (MDD) (Paquette et al., 2009).

The development and advances in simultaneous EEG-fMRI technique (e.g. Mulert and Lemieux, 2010), in which a subject wears an EEG cap inside an MRI scanner and EEG recordings are performed concurrently with fMRI data acquisition, has opened up new possibilities for neurofeedback research. Simultaneous EEG-fMRI provides the following important opportunities in the context of brain neuromodulation. First, electrophysiological correlates of rtfMRI-nf can be explored using EEG data recorded simultaneously with rtfMRI-nf training. Second, performance of EEG-nf can be validated based on fMRI data acquired simultaneously with EEG-nf training. Third, rtfMRI-nf can be dynamically modified using the simultaneously measured EEG activity. Finally, simultaneous multimodal rtfMRI-EEG neurofeedback can be provided to a subject to enable simultaneous self-regulation of both hemodynamic (rtfMRI) and electrophysiological (EEG) brain activity.

Here we report the first implementation of simultaneous multimodal rtfMRI-EEG neurofeedback (rtfMRI-EEG-nf) and its proof-of-concept application in training of emotional self-regulation. Our implementation of rtfMRI-EEG-nf is based on a novel, first-of-its-kind real-time integration of rtfMRI and EEG data streams for the purpose of brain neuromodulation.

During the experiment, healthy volunteers performed a positive emotion induction task by evoking happy autobiographical memories while simultaneously trying to regulate and raise two neurofeedback bars (rtfMRI-nf and EEG-nf) on the screen. The rtfMRI-nf was based on BOLD activation in a left amygdala region-of-interest (ROI), similar to our previous study of emotional self-regulation that used only rtfMRI-nf (Zotev et al., 2011). The EEG-nf, provided simultaneously with the rtfMRI-nf, was based on frontal hemispheric (left-right) EEG power asymmetry in the high-beta (beta3, 21−30 Hz) EEG frequency band.

Frontal EEG asymmetry is an important and widely used EEG characteristic of emotion and emotional reactivity (e.g. Davidson, 1992). It has been interpreted within the framework of the approach-withdrawal hypothesis (e.g. Davidson, 1992; Tomarken and Keener, 1998), which suggests that activation of the left frontal brain regions is associated with approach (i.e. higher responsivity to rewarding and positive stimuli), while activation of the right frontal regions is associated with withdrawal (i.e. tendency to avoid novel and potentially threatening stimuli). Brain activation is typically quantified by a reduction in alpha EEG power. The approach-withdrawal hypothesis applies to both emotional trait properties and emotional state changes in response to stimuli (e.g. Davidson et al., 1990; Sutton and Davidson, 1997; Coan and Allen, 2004). Numerous EEG studies have indicated that depression and anxiety are associated with reduced relative activation of the left frontal regions and increased relative activation of the right frontal regions (e.g. Tomarken and Keener, 1998; Thibodeau et al., 2006). Thus, frontal EEG power asymmetry is a natural target measure for EEG-nf aimed at training of emotional self-regulation, particularly in MDD patients.

Two studies have previously employed EEG-nf paradigms involving frontal EEG asymmetry. Allen at al. (2001) used EEG-nf based on the frontal EEG asymmetry in the alpha band for a group of healthy participants (Allen et al., 2001). They observed systematic changes in the asymmetry as the training progressed and associated changes in self-reported emotional responses. Paquette et al. (2009) applied EEG-nf based on EEG power in the high-beta band measured at two frontal and two temporal sites and used it in combination with psychotherapy sessions for a group of MDD patients. They reported a significant reduction in MDD symptoms associated with a significant decrease in high-beta EEG activity within the right frontal and limbic regions. This work followed up on the results of an earlier study (Pizzagalli et al., 2002) that demonstrated that MDD patients exhibited significantly higher resting EEG activity in the right frontal brain regions than healthy controls specifically in the high-beta band. The psychoneurotherapy (Paquette et al., 2009) led to significant changes in the high-beta EEG power asymmetry between the corresponding brain regions on the left and on the right.

In the present work, we implemented the EEG-nf based on the frontal EEG asymmetry in the high-beta band (21–30 Hz) rather than in the alpha band (8–13 Hz) because EEG-fMRI artifacts, caused by cardioballistic (CB) head motions as well as random head movements, are substantially reduced in this case. Also, electrophysiological activity in the high-beta band is relevant to depression, as mentioned above. The rtfMRI-EEG-nf was used in the present study for simultaneous upregulation of BOLD fMRI activation in the left amygdala ROI and frontal EEG power asymmetry in the high-beta band during the positive emotion induction task.

## Methods

### Integration of simultaneous rtfMRI and EEG data

Our implementation of rtfMRI-EEG-nf is based on a novel, first-of-its-kind real-time system, integrating simultaneous rtfMRI and EEG data streams. The system is designed for operation with a General Electric Discovery MR750 whole-body 3 T MRI scanner and a 128-channel MR-compatible EEG system from Brain Products GmbH. It represents a further development of the custom real-time MRI system (Bodurka and

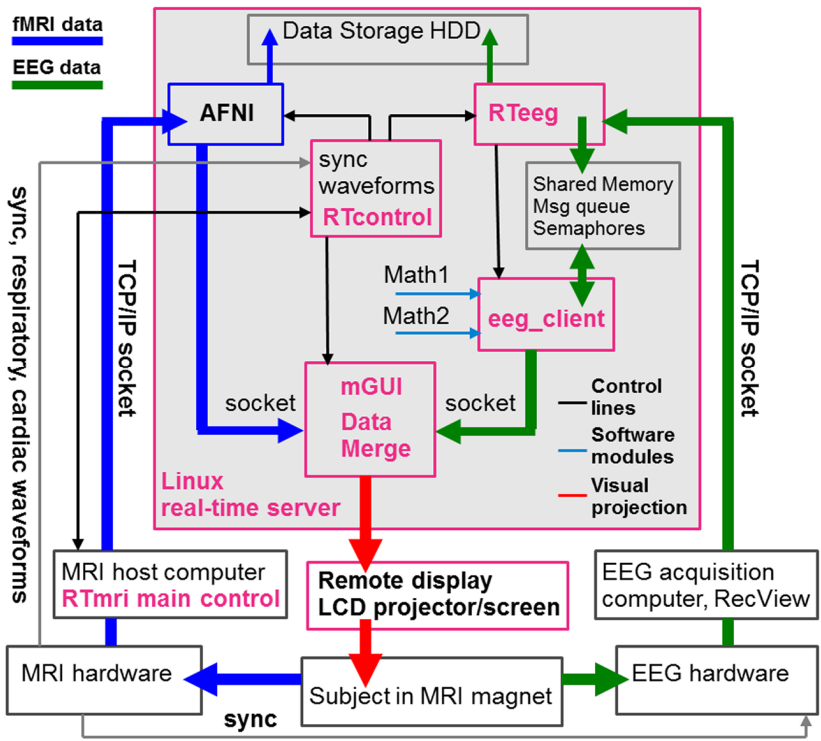

**Figure 1.** Block diagram of the real-time system for simultaneous rtfMRI-EEG neurofeedback. The diagram illustrates processing of simultaneous rtfMRI (blue arrows) and EEG (green arrows) data streams and their real-time integration into the rtfMRI-EEG neurofeedback data stream (red arrows).

Bandettini, 2008). A block diagram of the system is shown in Fig. 1.

The system design utilizes real-time features of AFNI (Cox, 1996; Cox and Hyde, 1997) and real-time functionality of BrainVision RecView software. The AFNI real-time plugin is used to perform real-time volume registration of fMRI data. It is also used to compute mean values of fMRI signals for several user-defined ROIs and export them for each fMRI volume in real time via a TCP/IP socket (Fig. 1). The RecView software makes it possible to partially remove MR and CB artifacts from the EEG data in real time using a built-in automated implementation of the average artifact subtraction method (Allen et al., 1998, 2000). RecView was custom modified to enable export of the corrected EEG data in real time through a TCP/IP socket (Fig. 1). Control and communication programs (shown in pink in Fig. 1) were written in Python (RTeeg, eeg_client, Math modules) and Perl (RTmri, RTcontrol, mGUI). The RTmri program runs on the MRI scanner's Linux control computer. The other programs run on a dedicated Linux real-time workstation with a kernel customized for high-speed inter-process communications with message queues, synchronization with semaphores, and large data exchange via shared memory.

The rtfMRI-nf signal is updated every *TR* and can be based on fMRI signal from a pre-selected ROI, such as the amygdala. It can also be computed using any combination of fMRI signals from multiple ROIs. Furthermore, our custom modification of the AFNI real-time plugin makes it possible to provide rtfMRI-nf based on real-time support vector machines (SVM) classification. The EEG-nf signal can be updated at a much faster rate (as often as every 100 ms). Real-time processing of the RecView-corrected data for EEG-nf is performed in Math modules (Fig. 1) utilizing NumPy functionality. It allows for: i) selection of an individual EEG channel or any combination of channels for analysis; ii) inspection of the EEG data and exclusion of data intervals with excessive artifacts; iii) real-time FFT spectrum analysis and computation of EEG power for any number of user-defined frequency bands; iv) calculation of any metrics for EEG-nf, such as the frontal EEG power asymmetry. The multimodal neurofeedback graphical user interface (mGUI, Fig. 1) integrates rtfMRI and EEG real-time data streams, computes the neurofeedback signals, and converts these signals into graphical representations viewed by the subject inside the scanner. mGUI is a multithreaded application supporting images, graphics primitives (such as bars), and text to form a dynamic display based on the rtfMRI-nf and EEG-nf signal levels (see Fig. 2a below).

Performance of the rtfMRI-EEG-nf system was extensively tested for multiple MRI/EEG hardware configurations prior to the system's use in human subject experiments. The tests utilized a standard MRI phantom and a specialized test signal generator for EEG. The system demonstrated robust real-time operation with 8- and 16-channel MRI head coil arrays and 32- and 128-channel EEG configurations.

*Experimental procedure*

The study was conducted at the Laureate Institute for Brain Research. The research protocol was approved by the Western Institutional Review Board (IRB). Six healthy subjects (mean age 24±9 years, four females) participated in the study. All the participants provided written informed consent as approved by the IRB. Each subject wore an EEG cap throughout the experiment. All the participants were neurofeedback-naïve.

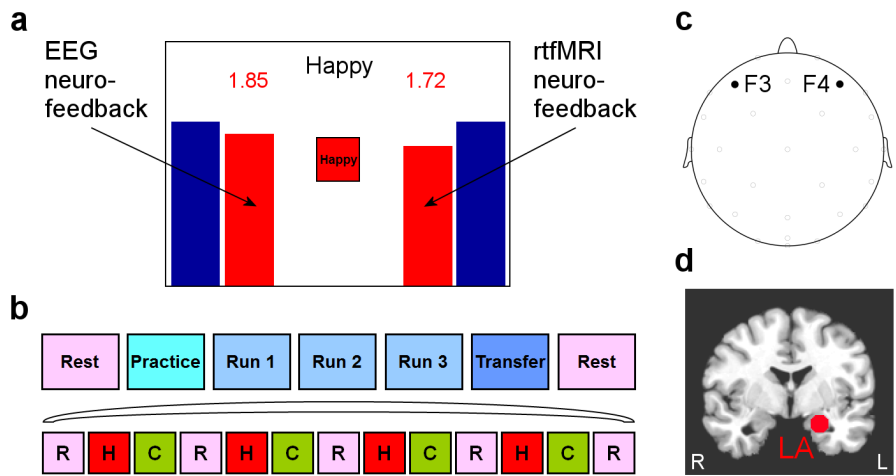

**Figure 2.** Elements of the experimental procedure. a) Neurofeedback GUI screen with red neurofeedback bars (EEG on the left, rtfMRI on the right) and blue target bars. b) Experimental protocol includes seven runs (Rest, Practice, Run 1, Run 2, Run 3, Transfer, and Rest), with each run (except Rest) consisting of 40 s long blocks of Happy Memories (H), Count (C), and Rest (R) conditions. c) Frontal EEG electrodes used to provide the EEG neurofeedback (F3 on the left, F4 on the right). d) Left amygdala (LA) ROI used to provide the rtfMRI neurofeedback.

The experimental procedure was developed based on the results of our previous work, which demonstrated that healthy subjects could learn to upregulate their left amygdala activation using rtfMRI-nf during a positive emotion induction task based on retrieval of happy autobiographical memories (Zotev et al., 2011). The main contribution of the present study is a proof-of-concept demonstration of rtfMRI-EEG-nf. Accordingly, each subject was presented with a neurofeedback display screen showing two neurofeedback bars: the rtfMRI-nf bar on the right side of the screen and the EEG-nf bar on the left (Fig. 2a).

The height of the rtfMRI-nf bar represented BOLD fMRI activation (with respect to a resting baseline) in the left amygdala (LA) ROI shown in Fig. 2d. The LA ROI was defined as a sphere of 7 mm radius centered at (−21, −5, −16) in the Talairach space (Talairach and Tournoux, 1988), as used in our previous work (Zotev et al., 2011). The rtfMRI-nf bar height was updated every 2 s.

The height of the EEG-nf bar represented a change in the frontal EEG power asymmetry (with respect to a resting baseline) between EEG electrodes F3 (on the left) and F4 (on the right) as depicted in Fig. 2c. The relative asymmetry was defined at each time point as $A=[P(F3)−P(F4)]/[P(F3)+P(F4)]$, where $P$ is the EEG power for a given channel in the high-beta (21−30 Hz) band. With this definition, an increase in the EEG-nf bar height corresponded to a reduction in high-beta power for the right frontal regions and/or enhancement in high-beta power for the left frontal regions (see *Discussion*). For offline statistical analysis, $A$ was normalized using Fisher transform as $An=\text{atanh}(A)$, which in this case reduces to $[\ln(P(F3))−\ln(P(F4))]/2$. Because $\ln(P)$ is commonly used to transform EEG power toward normal distribution (Gasser et al., 1982), $An$ is proportional to the difference in normalized power values for channels F3 and F4. The EEG-nf bar height was updated every 0.4 s. By controlling the rtfMRI-nf and EEG-nf bars, the participants learned to self-regulate both hemodynamic (rtfMRI) and electrophysiological (EEG) processes in their own brain simultaneously in real time.

The rtfMRI-EEG-nf training paradigm included three conditions: Happy Memories, Count, and Rest. The condition blocks are denoted by letters H, C, and R, respectively, in Fig. 2b. During the Happy Memories condition involving neurofeedback, the cue "Happy", two red neurofeedback bars (EEG-nf on the left, rtfMRI-nf on the right), and two blue target bars were displayed on the screen (Fig. 2a). The participants were instructed to feel happy by evoking happy autobiographical memories while also attempting to simultaneously increase the levels of both red bars on the screen toward the fixed levels of the blue target bars. During the Count condition, the subjects were shown the cue "Count" and a specific instruction to count backwards from 300 by subtracting a given integer. During the Rest condition, the participants were presented with the cue "Rest" and were asked to relax and breathe regularly while looking at the display screen. No bars were displayed during the Count and Rest conditions. Similarly, no bars were shown for the Happy Memories condition during the Transfer run (see below) and the instruction cue read "As Happy as possible".

The rtfMRI-EEG-nf experiment consisted of seven runs each lasting 8 min 40 s (Fig. 2b). The experiment began and ended with Rest runs (RE), during which the participants were instructed to let their minds wander while fixating on the display screen. During the Practice run (PR), the subjects were given an opportunity to become comfortable with the neurofeedback procedure. During the subsequent three training runs – Run 1 (R1), Run 2 (R2), and Run 3 (R3) – the participants underwent the neurofeedback training as instructed in detail before the experiment. These four runs consisted of alternating blocks of Rest (5 blocks), Happy Memories (4 blocks), and Count (4 blocks) conditions, each lasting 40 s (Fig. 2b). The target level for rtfMRI activation (blue bar on the right) was set to 0.5%, 1.0%, 1.5%, and 2.0% for the Practice run, Run 1, Run 2, and Run 3, respectively. The target level for the EEG asymmetry change (×10) was set, respectively, to 0.5, 1.0, 1.5, and 2.0 for the same four runs. During the Transfer run (TR), the participants were instructed to perform the same emotion induction task as during the neurofeedback training, but no neurofeedback information was provided and no bars were shown for the blocks of Happy Memories condition. The Count condition involved counting back from 300 by subtracting 3, 4, 6, 7, and 9 for the Practice run, Run 1, Run 2, Run 3, and the Transfer run, respectively. For more details of the experimental protocol and instructions given to the subjects refer to our previous work (Zotev et al., 2011).

*Data acquisition*

All functional and structural MR images were acquired using the General Electric MR750 3 T MRI scanner with a standard 8-channel receive-only head coil array. A single-shot gradient-recalled EPI sequence with Sensitivity Encoding (SENSE, Pruessmann et al., 1999) was employed for fMRI. To enable accurate correction of MR artifacts in EEG data, acquired simultaneously with fMRI, the EPI sequence was custom modified to ensure that the repetition time $TR$ was exactly 2000 ms (with 1 µs accuracy). The following EPI imaging parameters were used: FOV=240 mm, slice thickness=2.9 mm, slice gap=0.5 mm, 34 axial slices per volume, 64×64 acquisition matrix, echo time $TE$=30 ms, SENSE acceleration factor $R$=2, flip angle=90°, sampling bandwidth=250 kHz. The fMRI run time was 8 min 40 s. Three EPI volumes (6 s) were added at the beginning of the run to allow the fMRI signal to reach steady state and were excluded from data analysis. The fMRI voxel size was 3.75×3.75×2.9 mm3. We selected the EPI sequence with 64×64 acquisition matrix because it generated lower MR artifacts in simultaneously recorded EEG data than the higher-resolution sequence with 96×96 acquisition matrix used in our previous study (Zotev et al., 2011). Physiological pulse oximetry and respiration waveforms were recorded with 20 ms sampling interval simultaneously with fMRI. A photoplethysmograph placed on the subject's finger was used for pulse oximetry, and a pneumatic respiration belt was used for respiration measurements. A T1-weighted magnetization-prepared rapid gradient-echo (MPRAGE) sequence with SENSE was used to provide an anatomical reference for the fMRI analysis. It had the following parameters: FOV=240 mm, 128 axial slices per slab, slice thickness=1.2 mm, 256×256 image matrix, $TR$/$TE$=5.0/1.9 ms, SENSE factor $R$=2, flip angle=10°, delay time $TD$=1400 ms, inversion time $TI$=725 ms, sampling bandwidth=31.2 kHz, scan time=4 min 58 s.

The EEG recordings were performed simultaneously with fMRI using the Brain Products' MR-compatible EEG system configured for 32-channel operation. Each subject wore an MR-compatible EEG cap (BrainCap MR from EASYCAP GmbH) throughout the experiment. The cap is fitted with 32 EEG electrodes (including Ref), arranged according to the international 10-20 system, and one ECG electrode placed on the subject's back. To reduce head motions, two foam pads were inserted in the MRI head coil on both sides of the subject's head. The EEG amplifier (BrainAmp MR plus from Brain Products GmbH) was positioned just outside the MRI scanner bore near the axis of the magnet approximately 1 m away from the subject's head. The electrical cable connecting the EEG cap to the amplifier was fixed in place using sandbags. The amplifier was connected to the PC interface outside the scanner room via a fiber optic cable. The EEG system's clock was synchronized with the 10 MHz MRI scanner's clock using Brain Products' SyncBox device. The EEG signal acquisition was performed using BrainVision Recorder with 16-bit analog-to-digital conversion and 5 kS/s sampling providing 0.2 ms temporal and 0.1 µV measurement resolution. The EEG signals were measured relative to the standard reference (FCz). They were hardware-filtered during the acquisition in the frequency band between 0.016 Hz (10 s time constant) and 250 Hz.

*Real-time data processing*

In the present rtfMRI-nf implementation, the AFNI real-time plugin was used for volume registration of EPI images and export of mean fMRI signal values for the LA ROI (Fig. 2d) in real time. The rtfMRI-nf signal for each Happy Memories condition was defined as a percent signal change relative to the baseline obtained by averaging the rtfMRI signal for the preceding 40-second long Rest condition block. This neurofeedback signal (percent signal change) was updated every 2 s and displayed as the red bar on the right side of the mGUI screen (Fig. 2a). To reduce bar fluctuations due to noise in the fMRI signal, the bar height was computed at every time point as a moving average of the current and two preceding fMRI percent signal change values (Zotev et al., 2011).

For the EEG-nf, the RecView software was used to perform partial removal of MR and CB artifacts from the 32-channel EEG data in real time. The corrected data, downsampled to 250 S/s sampling rate (4 ms sampling interval), were exported in real time as data blocks of 8 ms duration (two data points per block for all channels). FFT power spectrum for the frontal EEG channels F3 and F4 (Fig. 2c) with FCz reference was computed every 0.4 s using a moving data interval of 2.048 s duration with Hann window. The relative EEG power asymmetry $A$ for F3 and F4 was calculated for the high-beta band as defined above (see *Experimental procedure* section). The change in $A$ for each Happy Memories condition was determined with respect to the baseline obtained by averaging $A$ values for the preceding 40-second long Rest condition block. This asymmetry change value (multiplied by 10) was updated every 0.4 s and displayed as the red bar on the left side of the mGUI screen (Fig. 2a). Similar to the rtfMRI-nf bar, the EEG-nf bar height was computed at every time point as a moving average of the current and two preceding asymmetry change values to reduce the bar fluctuations due to EEG noise.

*fMRI data analysis*

Offline analysis of the fMRI data was performed in AFNI. Pre-processing of single-subject fMRI data included correction of cardiorespiratory artifacts using AFNI implementation of the RETROICOR method (Glover et al., 2000). The cardiac and respiratory waveforms recorded simultaneously during each fMRI run were used to generate the cardiac and respiratory phase time series for the RETROICOR. Further fMRI pre-processing included slice timing correction and volume registration for all EPI volumes in a given run. Standard GLM analysis was then applied for each of the seven fMRI runs (Fig. 2b). The following regressors were included in the GLM model: two block-stimulus condition terms (Happy Memories, Count), six motion parameters as nuisance covariates, and five polynomial terms for modeling the baseline. The stimulus conditions for all runs (including the Rest) consisted of 40-second long blocks as defined in Fig. 2b. Hemodynamic response amplitudes were estimated using the standard regressors, constructed by convolving a boxcar function (representing the block duration) with the hemodynamic response function (HRF) using standard AFNI parameters. The GLM $\beta$ coefficients were computed for each voxel using the 3dDeconvolve AFNI program and then converted to percent signal changes for Happy vs Rest, Count vs Rest, and Happy vs Count contrasts. The resulting fMRI percent signal change maps for each run were spatially transformed to the Talairach space and re-sampled to 2×2×2 mm$^3$ isotropic voxel size. The voxel-wise percent signal change data were averaged within the LA ROI and used as a GLM-based measure of fMRI activation.

*EEG data analysis*

Offline analysis of the EEG data, acquired simultaneously with fMRI, was performed using BrainVision Analyzer 2. Removal of the MR and CB artifacts was based on the average artifact subtraction method, implemented in Analyzer 2. After the MR artifact removal, the EEG data were downsampled to 250 S/s sampling rate (4 ms sampling interval) and low-pass filtered at 80 Hz (48 dB/octave). The fMRI slice selection frequency (17 Hz) and its first harmonic (34 Hz) were removed by bandpass filtering together with the power line frequency (60 Hz). Data intervals exhibiting effects of significant head motions were carefully identified and excluded from further analysis. The CB artifact template was determined from the cardiac waveform recorded by the ECG channel, and the CB artifact to be subtracted was defined by a moving average over 21 cardiac periods. The standard reference (FCz) was used in EEG analysis because it yields more accurate results for the frontal electrodes than the common average reference (Hagemann et al., 2001).

Because this study focused primarily on the effects in the high-beta (21−30 Hz) EEG band, the EEG data were high-pass filtered at 15 Hz (48 dB/octave). Such filtering removed most of low-frequency artifacts and enabled a more accurate independent component analysis (ICA) in the frequency range of interest. The seven EEG runs were concatenated, and ICA was performed over the entire data length with exclusion of the motion-affected intervals. This approach ensured that independent components (ICs) corresponding to different artifacts were identified and removed in a consistent manner across all seven runs. The FastICA algorithm (Hyvärinen, 1999), implemented in Analyzer 2, was applied to the data from 31 EEG channels and yielded 31 ICs. Time course, spectrum, topography, and kurtosis value for each IC were carefully examined to identify and remove artifacts from the EEG data (see e.g. McMenamin et al., 2010).

Average EEG power spectra were computed for each experimental condition (Happy Memories, Count, and Rest) defined in Fig. 2b across each of the seven runs, including the Rest. A moving window FFT with 2.048 s data interval length, Hann window, and 50% interval overlap was applied after exclusion of the motion-affected intervals. The relative asymmetry $A$ in the high-beta band was then calculated for channels F3 and F4 and normalized as defined above (see *Experimental procedure*). The average $An$ values for Happy Memories and Rest conditions were then compared for each run.

*EEG-informed fMRI analysis*

To investigate possible BOLD fMRI correlates of the EEG-nf, we performed an EEG-informed fMRI data analysis (e.g. Mulert et al., 2010). A continuous wavelet transform with Morlet wavelets, available in Analyzer 2, was applied to the processed EEG data to compute EEG power as a function of time and frequency. The time resolution was 4 ms, and the frequency resolution was 0.5 Hz in the 0.5−30 Hz frequency range. Using these data, we calculated the normalized asymmetry $An$ for channels F3 and F4 in the high-beta band as a function of time. We then defined two regressors for inclusion in the psychophysiological interaction (PPI) analysis (Friston et al., 1997) of the fMRI data. One regressor was obtained by convolution of $An$ (converted to $z$-scores) with the HRF defined with 4 ms temporal resolution. For the other regressor, $An$ (converted to $z$-scores) was first multiplied by the contrast function (equal to +1 for Happy Memories, −1 for Count, and 0 for Rest conditions), and then convolved with the HRF. Both regressors were downsampled to $TR$ and included in the PPI analysis within the GLM framework. The first term described correlation of the EEG-asymmetry-based regressor with fMRI data. The second term described interaction [EEG-asymmetry-based regressor] × [Happy − Count], i.e. context-dependent correlation corresponding to Happy Memories vs Count condition contrast. We selected this contrast, because fMRI activation of the left amygdala and many other regions involved in emotion regulation increased during the Happy Memories condition blocks and decreased during the Count condition blocks, yielding a significant Happy−Count fMRI activation contrast (Zotev et al., 2011). The GLM model for each run also included two block-stimulus condition terms (Happy Memories, Count), six fMRI motion parameters (together with the same parameters shifted by one $TR$), five polynomial terms for modeling the baseline, and two covariates representing fMRI signals from white matter and CSF. The white matter covariate was computed as time course of mean fMRI signal from two spherical ROIs defined bilaterally within deep white matter. Similarly, the CSF covariate was calculated as time course of mean fMRI signal from two spherical ROIs defined within left and right ventricles. The PPI analysis was performed for each run using the 3dDeconvolve AFNI program. GLM-based $R^2$-statistics and $t$-statistics for the PPI interaction term were used to obtain correlation values $r$. The resulting maps were transformed to the Talairach space, re-sampled to 2×2×2 mm$^3$ isotropic voxel size, spatially smoothed (6 mm FWHM), and normalized using Fisher $r$-to-$z$ transform. Group $t$-test with respect to zero level was applied to the resulting single-subject maps to evaluate significance of the PPI interaction. Similar PPI analyses were also conducted using regressors based on normalized powers, $\ln(P(F3))$ and $\ln(P(F4))$, for the two individual channels.

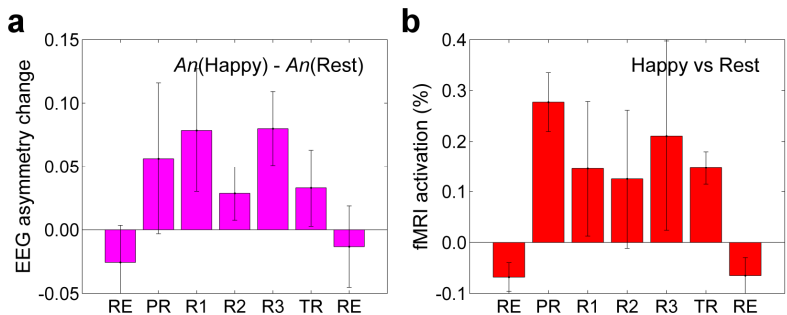

**Figure 3.** Effects of the simultaneous rtfMRI-EEG neurofeedback training for the group of six subjects (mean±SEM). a) Changes in normalized frontal EEG power asymmetry $An$ in the high-beta band (21-30 Hz) for the Happy Memories condition with respect to the Rest condition. b) fMRI activation of the left amygdala ROI for the Happy Memories condition with respect to the Rest condition.

*Effects of EEG artifacts*

To evaluate effects of different EEG artifacts on the measures of EEG asymmetry, we denoted successive stages of the offline EEG signal processing performed in the present work as Steps 1, 2, and 3. Step 1 involves the partial removal of MR and CB artifacts using the average artifact subtraction method and exclusion of data intervals affected by significant head motions. This step is similar to the real-time processing of the EEG data. The subsequent removal of residual MR and CB artifacts based on the offline ICA is denoted as Step 2. Finally, Step 3 includes the removal of muscle and rapid eye movement (saccadic) artifacts identified in the same ICA analysis. To estimate power contributions of different artifacts, we computed average power values in the high-beta band separately for two channels (F3 and F4) and two conditions relevant to the EEG-nf (Happy Memories and Rest) after Step 1, Step 2, and Step 3 for each run. The average power for each channel/condition combination after Step 1 was set to 100%. Comparison of the average power values (for each channel/condition combination) after Step 2 and after Step 1 makes it possible to estimate residual CB and MR artifact contributions. Comparison of the average power values after Step 3 and after Step 2 allows estimation of muscle artifact contributions. Finally, % contribution estimates for each artifact type were averaged over the four channel/condition combinations, across all runs, and across all subjects.

**Results**

The rtfMRI-EEG-nf system has demonstrated safe, stable, and reliable real-time performance in the described experiments utilizing the 8-channel MRI head coil array and the 32-channel MR-compatible EEG caps. During each experiment, the rtfMRI-nf and EEG-nf signal values used for the visual neurofeedback display are incrementally saved to data files. These values have invariably shown close agreement with the corresponding

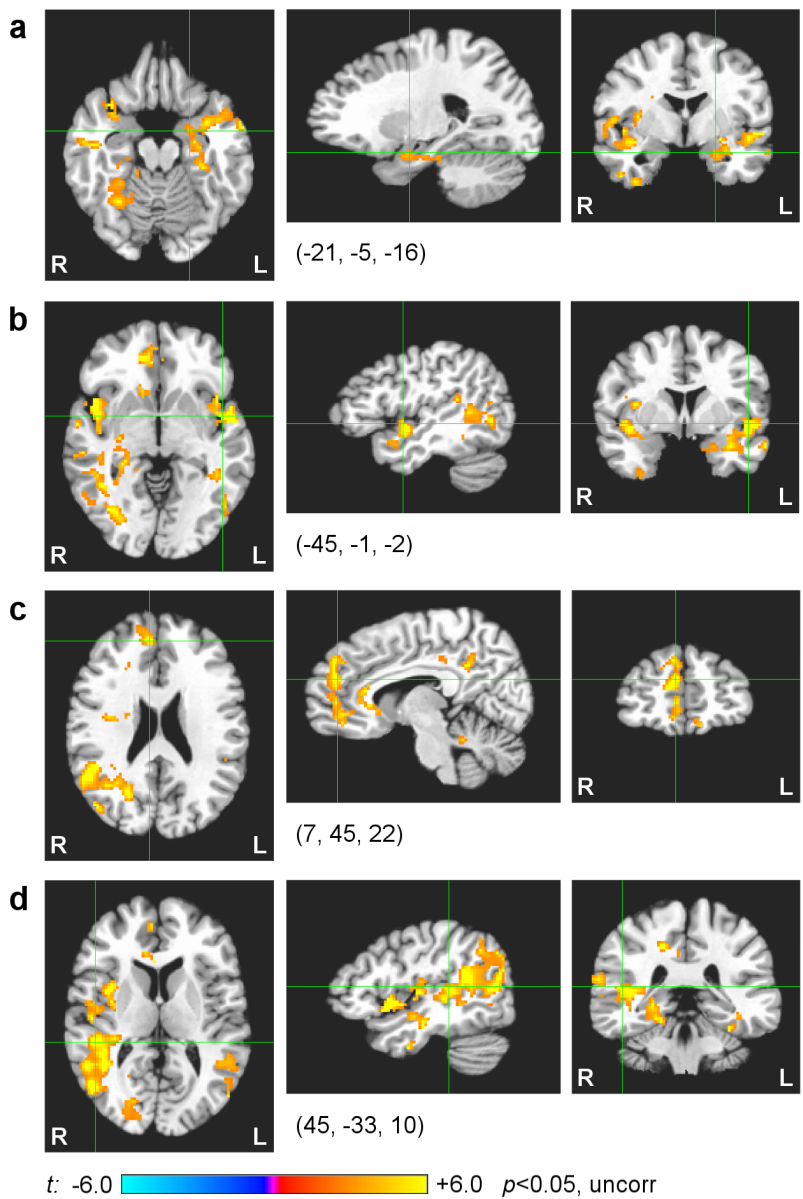

**Figure 4.** Results of the EEG-informed analysis of the fMRI data. The figures show group statistical maps for the psychophysiological interaction [EEG-asymmetry-based regressor] × [Happy − Count]. The maps are projected onto the standard anatomical template (TT_N27) in the Talairach space. Coordinates of the point marked by the green crosshaires are specified underneath each set of figures. a) Cluster including the left amygdala. b) The same cluster also includes the left insula. c) Cluster including the right dorsomedial prefrontal cortex and anterior cingulate. d) Cluster including the right superior temporal gyrus. See Table 1 for details.

fMRI and EEG characteristics obtained in offline analyses of the same fMRI and EEG data.

Results of the proof-of-concept rtfMRI-EEG-nf experiment are exhibited in Fig. 3. Each bar in Fig. 3a represents the average difference in the normalized relative asymmetry $A_n$ between the Happy Memories and Rest conditions for a given run averaged across the group of six subjects. The error bars are standard errors of the means (SEM). For the Rest runs, the blocks of Happy Memories and Rest conditions were defined in the same way as for the other runs (Fig. 2b). Similarly, each bar in Fig. 3b represents a mean fMRI percent signal change for the LA ROI, averaged for Happy Memories conditions in a given run and across the group. The mean LA ROI results for each participant were obtained from the GLM analysis using the same stimulus regressors for each run, including the Rest runs. The individual fMRI activation results for the six subjects are exhibited in Supplementary Fig. 1, and the individual EEG asymmetry change results are shown in Supplementary Fig. 2.

The results in Fig. 3 demonstrate that the participants were able to increase their average BOLD fMRI activation in the left amygdala region, as well as their frontal EEG asymmetry in the high-beta band during the neurofeedback runs (Practice, Run 1, Run 2, Run 3) and the Transfer run. Because of the limited number of subjects (six) in this proof-of-concept study, the group results in Fig. 3 are statistically significant only for selected runs. The EEG asymmetry change is significant for Run 3 ($t(5)=2.75$, $p<0.04$). The LA fMRI activation is significant for the Practice run ($t(5)=4.81$, $p<0.005$) and for the Transfer run ($t(5)=4.51$, $p<0.006$). Results of higher significance can be expected for a larger experimental group (see Zotev et al., 2011).

Results of the psychophysiological interaction (PPI) analysis using regressors based on the EEG asymmetry $A_n$ are shown in Fig. 4. The analysis was performed as described above (see *EEG-informed fMRI analysis* section). Results of only one neurofeedback run (among Practice, Run 1, Run 2, and Run 3), characterized by the highest mean Happy vs Rest LA ROI activation, were included in the PPI analysis from each participant. The group statistical maps for the [EEG-asymmetry-based regressor] × [Happy−Count] interaction term in Fig. 4 are thresholded at $p<0.05$ (uncorrected). Significant positive interaction effects are observed in Fig. 4 for several regions involved in emotion regulation and visual processing. Because of the limited number of subjects in this proof-of-concept work, only the largest interaction clusters (cluster size > 200 voxels, i.e. 1.6 cm$^3$) are shown in the figure. Their characteristics are detailed in Table 1. It should be noted that the cluster with the highest t-score in the left insula region (BA 22, Fig. 4b) also includes parts of the left amygdala and the center of the LA ROI (Fig. 4a). The cluster with the interaction peak $t$-value in the right

Table 1. Results of the EEG-informed analysis of the fMRI data. Group statistical data for the [EEG-asymmetry-based regressor] × [Happy − Count] psychophysiological interaction. Location of the point with the maximum group *t*-score and the number of voxels are specified for each cluster.

| Region | Laterality | Talairach coordinates | | | Size | *t*-score |
|---|---|---|---|---|---|---|
| | | x | y | z | | |
| Superior temporal gyrus (BA 41) | R | 45 | −33 | 10 | 3041 | 10.99 |
| Orbitofrontal cortex (BA 47) | R | 45 | 13 | −6 | 1098 | 7.16 |
| Insula (BA 22) | L | −45 | −1 | −2 | 948 | 13.46 |
| Dorsomedial prefrontal cortex (BA 9) | R | 7 | 45 | 22 | 780 | 10.11 |
| Middle temporal gyrus (BA 39) | L | −49 | −67 | 12 | 410 | 3.39 |
| Lingual gyrus | R | 25 | −75 | 2 | 310 | 9.84 |
| Posterior cingulate cortex (BA 31) | R | 17 | −35 | 38 | 293 | 7.68 |
| Inferior temporal gyrus (BA 20) | R | 45 | −5 | −32 | 246 | 4.37 |

BA − Brodmann areas; L − left; R – right; Size – cluster size, minimum 200 voxels.
*$p < 0.05$, uncorrected.

orbitofrontal cortex (BA 47) also includes portions of the right insula (BA 13), as can be seen in Fig. 4b. The cluster with the highest t-score in the right dorsomedial prefrontal cortex (BA 9, Fig. 4c) also comprises parts of the right anterior cingulate (BA 24/32, Fig. 4c). Implications of these PPI analysis results are discussed in detail below.

In addition to the PPI analysis for the EEG asymmetry, we conducted similar PPI analyses using signals from the two individual channels, and compared group average values of [F3-power-based regressor] × [Happy−Count] and [F4-power-based regressor] × [Happy−Count] interaction terms at different locations, including the center of the LA ROI and the points specified in Table 1. The average interaction values corresponding to F3 and F4 had opposite signs at the center of the LA ROI (+0.034 for F3 vs −0.006 for F4), at the left insula (+0.009 vs −0.024), and at the superior temporal gyrus (+0.012 vs −0.028). For the other locations in Table 1, the average interaction values had the same signs for F3 and F4, but were invariably more positive for F3 than for F4.

Fig. 5 illustrates dependence of the EEG asymmetry results on the presence of different artifacts in the EEG data. The analysis was performed as described above (see *Effects of EEG artifacts* section). According to Fig. 5a, the positive changes in the average normalized EEG asymmetry *An* for the high-beta band are already observed after the basic EEG processing (Step 1). They become larger after Step 2 (Fig. 5b). The average *An* changes can either increase or decrease after Step 3, but remain positive (Fig. 5c, the same as Fig. 3a). The fact that the positive asymmetry changes for the Happy Memories conditions relative to the Rest conditions generally become more pronounced as more artifacts are removed from the EEG data indicates that these changes cannot be attributed entirely to the EEG artifacts.

The results in Fig. 5d suggest that EEG artifacts contribute substantially to the average EEG power in the high-beta band for channels F3 and F4 after the basic EEG processing (Step 1). These are predominantly residual MR and CB artifacts (~50%). Notably, the residual CB artifacts provide most substantial contributions when the average artifact subtraction procedure fails to correct the CB artifacts due to major variations in the cardiac waveform profile. Muscle artifacts, which have inherently broad spectra, also contribute to the average EEG power for channels F3 and F4 (~20%). Because only the basic EEG processing (and not ICA) can presently be performed in real time, the residual MR, residual CB, and muscle artifacts present a serious challenge for potential applications of rtfMRI-EEG-nf, as discussed below.

## Discussion

We developed a novel system for real-time integration of simultaneous rtfMRI and EEG data streams (Fig. 1) and used it to implement, for the first time, simultaneous multimodal rtfMRI and EEG neurofeedback (Fig. 2). We demonstrated that healthy participants were able to simultaneously upregulate their frontal high-beta EEG asymmetry and left amygdala BOLD fMRI activation (Fig. 3) using the rtfMRI-EEG-nf during the positive emotion induction task.

The regulation of asymmetric electrophysiological activity of the frontal brain regions in either the alpha (8−13 Hz) or the high-beta (21−30 Hz) band by means of EEG-nf has been associated with changes in emotional state (Allen et al., 2001; Paquette et al., 2009). In the present work, we defined the frontal high-beta EEG asymmetry as left-to-right asymmetry, $An \sim \log(P(\text{left})) - \log(P(\text{right}))$, where *P* is the EEG power in the high-beta band. This EEG band selection and asymmetry definition are based on empirical evidence. It has been demonstrated using brain electrical tomography that MDD patients exhibit significantly more high-beta EEG power in the right superior and inferior frontal brain regions than healthy subjects (Pizzagalli et al., 2002). Thus, the left-to-right frontal asymmetry in the high-beta band is lower in MDD patients than in healthy controls. More positive values of the left-to-right high-beta asymmetry are associated with reduced severity of depressive symptoms in melancholic MDD patients (Pizzagalli et al., 2002). It has also been suggested that stronger high-beta power can be interpreted as an indication of increased excitatory brain activity (Pizzagalli et al., 2002). The study by Paquette et al. (2009) further demonstrated that reduction in the high-beta EEG power within right anterior brain regions as a result of the psychoneurotherapy significantly correlated with reduction in MDD symptoms. It should be noted that the high-beta EEG-nf paradigm in that study did not include a single specific target measure for EEG-nf, and did not provide a specific strategy beyond general reduction in negative thoughts and feelings (Paquette et al., 2009). Nevertheless, the EEG-nf training in combination with psychotherapy led to significant positive changes in the left-to-right EEG asymmetry in the high-beta band for the corresponding brain regions on the left and on the right. Another relevant study examined selective attention to angry facial expressions in healthy subjects (Schutter et al., 2001). The study found that reduced left-to-right EEG asymmetry in the beta (13−30 Hz) band significantly correlated with more avoidant reaction to angry faces. Avoidance of angry facial expressions is associated with affective withdrawal and elevated depressive symptoms (Schutter et al., 2001). Taken together, these results suggest that voluntary upregulation of left-to-right frontal EEG asymmetry in the high-beta band may have a therapeutic effect of reducing negative emotions and severity of MDD symptoms.

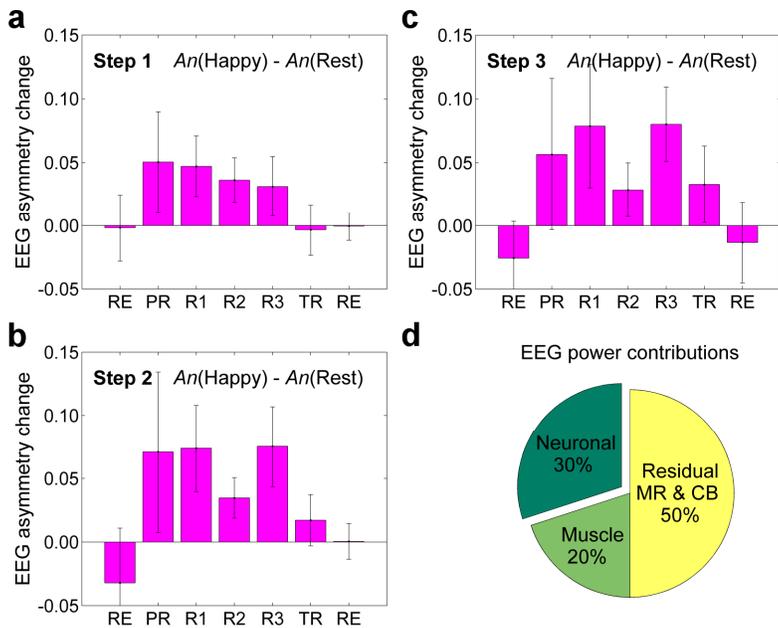

**Figure 5.** Effects of successive EEG signal processing steps on the resulting measures of the EEG asymmetry change in the high-beta (21-30 Hz) band. a) Partial removal of magnetic resonance (MR) and cardioballistic (CB) artifacts using the average artifact subtraction method and exclusion of motion-affected intervals (Step 1). b) ICA-based removal of residual MR and CB artifacts (Step 2). c) ICA-based removal of muscle and eye movement artifacts (Step 3). d) Contributions from different sources to the average EEG power in the high-beta band for channels F3 and F4 after Step 1.

We demonstrated in the present work that the EEG-nf based on the frontal high-beta EEG asymmetry can be naturally combined and used simultaneously with the rtfMRI-nf based on the amygdala activation. We recently showed that healthy volunteers are able to upregulate fMRI activation in their left amygdala using rtfMRI-nf during the positive emotion induction task (Zotev et al., 2011). Additionally, preliminary results of the same rtfMRI-nf procedure for MDD patients indicated that such rtfMRI-nf training of the amygdala is accompanied by decreases in state measures of depression (Young et al., 2012). Therefore, the two types of neurofeedback are generally compatible. Conceivably, the rtfMRI-EEG-nf may prove more efficient in training of emotional self-regulation than either the rtfMRI-nf or the EEG-nf applied separately.

The EEG-informed fMRI analysis (see *Results* section) revealed that several brain regions exhibited positive correlation with the high-beta EEG asymmetry that was significantly stronger during the Happy Memories condition than during the Count condition (Fig. 4 and Table 1). Importantly, the positive PPI interaction effect was observed in the left amygdala region, which was the target area for the rtfMRI-nf (Fig. 4a). This result suggests that the EEG-nf and the rtfMRI-nf employed in the present study are mutually compatible, i.e. increases in the frontal high-beta EEG asymmetry during the Happy Memories condition with rtfMRI-EEG-nf are associated with increases in fMRI activation of the left amygdala. Notably, the fact that this positive effect is only moderately significant at the group level (typically $0.01 < p < 0.05$ or $p \geq 0.05$ in the left amygdala region) indicates that the two types of neurofeedback are not mutually redundant. Several other anterior brain regions also showed significant correlations with the frontal high-beta EEG asymmetry for the same condition contrast (Table 1). They include the left and right insula (Fig. 4b), the right orbitofrontal cortex, the right dorsomedial prefrontal cortex and anterior cingulate (Fig. 4c). These regions play important roles in emotion processing and regulation. In particular, they exhibited significant correlations with the left amygdala during the rtfMRI-nf training in our previous study (Zotev et al., 2011). Posterior brain areas exhibiting significant PPI interaction effects include the right superior temporal gyrus (Fig. 4d) and the lingual gyrus (Table 1). The right superior temporal gyrus is involved in visual search (Ellison et al., 2004), as well as decision-making, particularly strategy changes based on the previous experience (Paulus et al., 2005). The lingual gyrus is involved in selective visual attention and target processing (Hopfinger et al., 2000). Engagement of these regions can be associated with processing of visual neurofeedback information. Another interesting result is the fact that the average PPI interaction values for the individual signals from channels F3 and F4 (rather than their asymmetry) have opposite signs at the considered locations within the left amygdala, the left insula, and the right superior temporal gyrus (see *Results* section). This observation suggests that an increase in fMRI activation of the left amygdala (or at least its part) during the Happy Memories conditions with rtfMRI-EEG-nf relative to the Count conditions is associated with an increase in high-beta EEG power on the left (F3) and a decrease in high-beta EEG power on the right (F4) for the same condition contrast. This is another important indication that the rtfMRI-nf and the EEG-nf are compatible and may have mutually consistent therapeutic effects.

In general, rtfMRI-EEG-nf can be expected to have advantages over rtfMRI-nf and EEG-nf used individually. First, rtfMRI-EEG-nf allows simultaneous multimodal regulation of both hemodynamic (BOLD) brain responses and their underlying electrophysiological processes, which are profoundly related yet exhibit important differences (Logothetis et al., 2001). Second, the high temporal resolution of EEG-nf makes it a valuable complement to rtfMRI-nf (with its relatively long *TR* and sluggish BOLD response) for those training paradigms, in which neuromodulation speed is essential. Third, rtfMRI-EEG-nf makes it possible to dynamically modify an experimental protocol (e.g. target levels for neurofeedback, sequence and durations of tasks) and an individual strategy in real time based on both rtfMRI-nf and EEG-nf information. Finally, the availability of both rtfMRI-nf and EEG-nf at any time during an experiment may lead to development of new training paradigms in which the effects of rtfMRI-nf could be approximated using only EEG-nf. This would have profound practical significance, because instrumentation for EEG-nf is considerably more affordable and portable than that for rtfMRI-nf.

The main challenge for practical implementation of EEG-nf with simultaneous fMRI (including rtfMRI) is real-time removal of EEG-fMRI artifacts. The average artifact subtraction method, implemented e.g. in BrainVision RecView software, is fairly efficient for removal of MR and CB artifacts provided that these artifacts remain stable over extended periods of time (tens of seconds). Any rapid changes in artifact properties cannot be corrected in this way, and result in substantial residual artifacts left in the EEG data. For example, even a small displacement of the EEG array with respect to the MRI scanner's isocenter due to head motion alters magnitude and spatial properties of MR artifacts, leading to residual MR artifacts. Similarly, any significant variation in the cardiac waveform profile leaves residual CB artifacts after the average CB artifact subtraction. More random artifacts, such as artifacts due to random head movements and muscle artifacts, are even more difficult to correct in real time. EEG data intervals, affected by drastic head movements, can be excluded, for example, by imposing thresholds on EEG signal's magnitude and its variation over a pre-selected time duration. Similarly, data intervals exhibiting severe muscle artifacts can be identified by setting thresholds for EEG power at higher frequencies (say, >30 Hz) for peripheral channels. However, a more stringent thresholding in either case would lead to exclusion of larger portions of EEG data, potentially preventing the EEG-nf signal from being a real-time measure of brain activity. To make matters worse, head motions, cardiac waveform variations, and muscle activity correlate with experimental tasks and tend to increase with increasing task difficulty. Residual MR and CB artifacts, as well as muscle, saccadic and other EEG artifacts, can be quite accurately removed in offline EEG analysis using ICA. However, no ICA can presently be performed on EEG data in real time.

In the present work, the EEG-nf was carefully designed to reduce the effects of EEG-fMRI artifacts on the EEG-nf signal as much as possible irrespective of the artifact removal procedure. First, the high-beta (21–30 Hz) EEG band was selected instead of the alpha or lower frequency bands. This greatly reduces the effects of CB and random motion artifacts on the EEG-nf signal. The high-beta band lies between the MR artifact spectral peaks at the fMRI slice excitation frequency (17 Hz for our EPI sequence) and its first harmonic (34 Hz). Second, the EPI sequence with lower (3.75×3.75 mm$^2$) in-plane resolution (64×64 acquisition matrix) was chosen instead of the readily available sequence with higher resolution (96×96 matrix as in Zotev et al., 2011). This reduces the maximum values of the imaging gradients and the corresponding MR artifacts. To improve the average MR artifact subtraction, the EEG system's clock was synchronized with the MRI scanner's clock. Third, the frontal EEG channels F3 and F4 were selected as active channels for the EEG-nf. These channels experience much lower muscle and saccadic artifacts than the peripheral EEG channels (such as Fp1, Fp2, F7, F8, T7, T8). They also exhibit lower CB and random motion artifacts (with FCz reference) than most of the EEG channels. These design considerations ensured that the EEG-nf signal was a reliable (though not perfectly accurate) measure of changes in frontal EEG power asymmetry after the basic real-time EEG data processing, as demonstrated by the results in Fig. 5a and Fig. 5c.

Despite the successful demonstration of the rtfMRI-EEG-nf in this proof-of-concept work, much remains to be done to improve the EEG-nf performance. According to Fig. 5d, residual MR and CB artifacts constitute as much as ~50% of the average EEG signal power for channels F3 and F4 in the high-beta band after the basic real-time EEG signal processing. Clearly, more efficient techniques for removal of EEG-fMRI artifacts in real time will have to be implemented. It should be noted that MR, CB, and random motion artifacts in EEG-fMRI have essentially the same physical mechanism and arise due to Faraday induction within spurious contours formed by scalp conductivity paths together with EEG electrode leads (see e.g. Zotev et al., 2012). A detailed analysis of such effects makes it possible to develop novel hardware and software solutions for more accurate artifact removal for EEG-fMRI in general (Zotev et al., 2012) and for rtfMRI-EEG-nf in particular. Our work in this direction is currently under way.

## Conclusion

We have demonstrated the feasibility of simultaneous self-regulation of both hemodynamic and electrophysiological activity of the human brain using the first-ever implementation of simultaneous multimodal rtfMRI and EEG neurofeedback. rtfMRI-EEG-nf is a novel neuromodulation modality with its own unique opportunities and challenges. Our results suggest potential applications of rtfMRI-EEG-nf in the development of novel cognitive neuroscience research paradigms and enhanced cognitive therapeutic approaches for major neuropsychiatric disorders, particularly depression.

## Acknowledgments

This work was supported by the Laureate Institute for Brain Research and the William K. Warren Foundation. We would like to thank Dr. Patrick Britz and Dr. Robert Störmer of Brain Products, GmbH for their continuous help and excellent technical support. We also thank Dr. Maria Schatt of Brain Products, GmbH for her kind help with custom modification of the BrainVision RecView software.